\documentclass[sigconf]{acmart}

\usepackage{color}
\usepackage[ruled,linesnumbered]{algorithm2e}
\SetKwInput{KwInput}{Input}               
\SetKwInput{KwOutput}{Output}
\SetKwInput{KwInit}{Init}
\usepackage{bm}
\usepackage{multirow}
\usepackage{array}
\usepackage{balance}
\usepackage{url}
\usepackage{multirow}
\usepackage{enumitem}
\usepackage{enumerate}
\usepackage{caption}
\usepackage{subcaption}
\usepackage{colortbl}
\usepackage{xcolor}

\AtBeginDocument{%
  \providecommand\BibTeX{{%
    \normalfont B\kern-0.5em{\scshape i\kern-0.25em b}\kern-0.8em\TeX}}}

\copyrightyear{2024}
\acmYear{2024}
\setcopyright{acmlicensed}
\acmConference[WWW '24 Companion] {Companion Proceedings of the ACM Web Conference 2024}{May 13--17, 2024}{Singapore, Singapore.}
\acmBooktitle{Companion Proceedings of the ACM Web Conference 2024 (WWW '24 Companion), May 13--17, 2024, Singapore, Singapore}
\acmISBN{979-8-4007-0172-6/24/05}
\acmDOI{10.1145/3589335.3651548}

\begin{document}

\title{Proactive Recommendation with Iterative Preference Guidance}


\author{Shuxian Bi$^{1}$, Wenjie Wang$^{2*}$, Hang Pan$^{1}$, Fuli Feng$^{1*}$ and Xiangnan He$^1$}
\def\authors{Shuxian Bi, Wenjie Wang, Hang Pan, Fuli Feng and Xiangnan He}
\affiliation{
\institution{$^1$University of Science and Technology of China,\\ $^2$National University of Singapore}
\country{}
}
\email{{shuxianbi, hungpaan}@mail.ustc.edu.cn, {wenjiewang96, fulifeng93, xiangnanhe}@gmail.com}
\thanks{$*$ Corresponding authors: Wenjie Wang and Fuli Feng. 

This work is supported by the National Key Research and Development Program of China (2022YFB3104701), the National Natural Science Foundation of China (62272437), and the CCCD Key Lab of Ministry of Culture and Tourism.}

\renewcommand{\shortauthors}{Shuxian Bi, Wenjie Wang, Hang Pan, Fuli Feng \& Xiangnan He}

\begin{abstract}



Recommender systems mainly tailor personalized recommendations according to user interests learned from user feedback. 
However, such recommender systems passively cater to user interests and even reinforce existing interests in the feedback loop, leading to problems like filter bubbles and opinion polarization. 
To counteract this, \textit{proactive recommendation} actively steers users towards developing new interests in a target item or topic by strategically modulating recommendation sequences.
Existing work for proactive recommendation faces significant hurdles:  
1) overlooking the user feedback in the guidance process; 
2) lacking explicit modeling of the guiding objective;
and 3) insufficient flexibility for integration into existing industrial recommender systems.
To address these issues, we introduce an \textbf{I}terative \textbf{P}reference \textbf{G}uidance (IPG) framework. IPG performs proactive recommendation in a flexible post-processing manner by ranking items according to their IPG scores that consider both interaction probability and guiding value. These scores are explicitly estimated with iteratively updated user representation that considers the most recent user interactions.
Extensive experiments validate that IPG can effectively guide user interests toward target interests with a reasonable trade-off in recommender accuracy. The code is available at \url{https://github.com/GabyUSTC/IPG-Rec}.
\end{abstract}




\maketitle

\section{Introduction}\label{sec:introduction}
Recommender Systems (RSs) play a crucial role in various Internet platforms such as e-commerce.
Traditionally, RSs continuously learn user interests from historical user feedback and provide items passively catering to the inferred interests.
Along the feedback loop\footnote{A cyclic process that user interests and behaviors are updated by the recommended items and RSs is self-reinforcing the updated data.} \cite{DBLP:conf/recsys/CurmeiHRH22, DBLP:conf/chi/CosleyLAKR03} in the recommendation scenario, such catering recommendation tends to build filter bubbles, narrowing user interests and causing detrimental social issues like opinion polarization \cite{DBLP:journals/corr/abs-1910-05274}.


\begin{figure}[!t]
\setlength{\abovecaptionskip}{0cm}
\setlength{\belowcaptionskip}{-0.3cm}
\centering
\includegraphics[scale=0.55]{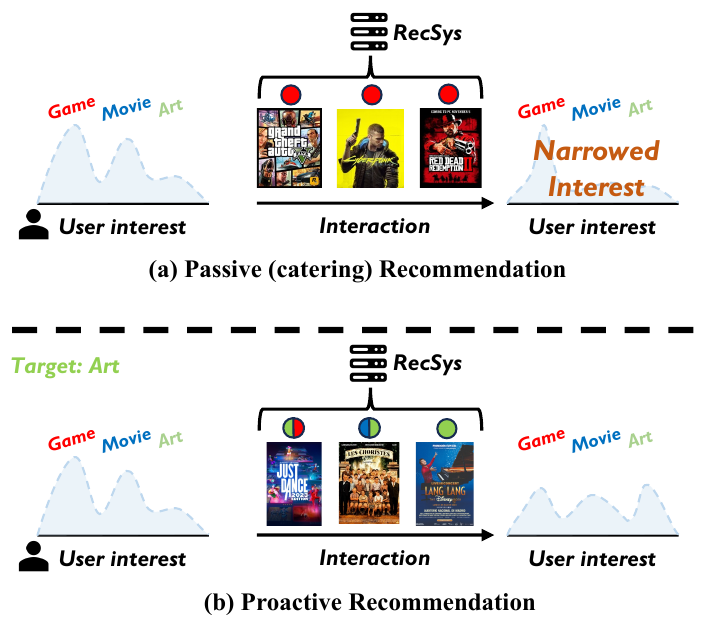}
\caption{Illustration of proactive recommendation. The color of the circle above each item indicates the topics.}
\vspace{-1mm}
\label{fig:target-oriented}
\end{figure}

To overcome the limitation of traditional RSs, \textit{proactive recommendation} actively guides users to jump out historical interests based on the assumption that recommended items can affect user interests~\cite{DBLP:conf/wsdm/CaiFWYLX23}.
As Figure \ref{fig:target-oriented} shows, proactive recommendation begins by selecting a target item or target topic representing the expected interest and aims to actively guide user interests towards this target by adjusting recommendation lists. 
The guiding process unfolds iteratively in a multi-round manner. During each round, RS recommends items\footnote{As an initial attempt, the number of recommendations in each round is set as one, which is the default choice in many scenarios such as micro-video recommendation.} to the user, triggering the evolution of user interests and interactions on the recommendation accordingly.

Influential Recommender System (IRS) \cite{DBLP:journals/corr/abs-2211-10002} represents an initial effort toward this objective. 
It encodes both historical user feedback and the target item to generate a sequence of items that bridge the current interest and the target item, termed an \textit{influential path}.
However, IRS faces three limitations that prompt a quest for a more effective guiding strategy:
\begin{itemize}[leftmargin=*]
    \item IRS assumes that users passively accept all recommended items so that generates unrealistic influence paths, which introduce a substantial disparity with real-world scenarios.
    \item The guiding objective is not explicitly manifested. IRS relies solely on the incorporation of an additional attention mask on the target item, without employing explicit guiding signals during both the training and inference stages. This implicit modeling approach may result in a suboptimal performance of its guiding strategy.
    \item IRS lacks of flexibility. The additional attention mask module brings challenges for integration with existing industrial RS.
\end{itemize}

To tackle these challenges, we propose an \textbf{I}terative \textbf{P}reference \textbf{G}uidance (\textbf{IPG}) framework for proactive recommendation.
To capture real-time user interest during the guidance process, IPG encodes the most recent user feedback as the user representation. This approach helps mitigate the issue of unrealistic influence paths observed in IRS where users may not follow the guidance due to the lack of accurate real-time user interest modeling.
Leveraging the captured user interests, IPG introduces an explicit IPG score, which consists of a guiding score and interaction probability, aiming to rank items based on the increase in user interest toward the target item. This explicit IPG score clarifies the objective of the guiding process, leading to improved performance in guidance compared to the implicit method IRS.
IPG is model-agnostic and introduces no additional training objectives or parameters. This characteristic renders IPG a flexible post-processing strategy that can be easily integrated into various sequential recommendation models.
Moreover, to gather user feedback in each iteration for evaluation, we design an interactive recommendation simulator that simulates users' preference evolution, boredom effect, and user feedback.

The main contributions of this work are summarized as follows:
\begin{itemize}[leftmargin=*]
    \item We emphasize the necessity of proactive recommendation and utilize real-time user feedback for iterative preference guidance.
    \item We propose the IPG framework for proactive recommendation, which is flexible and effective for sequential recommenders.
    \item We design an interactive recommendation simulator to verify the effectiveness of IPG. Experiments show the superiority of our proposal in preference guidance.
\end{itemize}

\section{Method}\label{method}


\subsection{Task Formulation}
Let $u \in \mathcal{U}$ and $i \in \mathcal{I}$ represent a user and an item, respectively. Each user corresponds to a historical interaction sequence $S_{u}^t = [i_1, i_2, \dots i_t]$.
After receiving and interacting with a new item, the user preference will change accordingly. Specifically, the preference of user $u$ to item $i$ at time-step $t$ is denoted as $p_{ui}^t$. 

Given a target item $j$, the goal of proactive recommendation is to improve the user preference on the target item within $l$ recommendation steps. Formally, the objective is maximizing the preference improvement $p_{uj}^{t+l} - p_{uj}^{t}$. Simultaneously, the recommendation path from time-step $t+1$ to $t+l$ should attract sufficient user interactions with respect to conventional metrics such as Hit Ratio. In other words, our target is to learn the recommendation strategy for the following $l$ recommendation steps with consideration of both guiding and interaction objectives.

\subsection{Iterative Preference Guidance}
Considering the flexibility, we focus on achieving proactive recommendation based on a well-trained sequential recommendation model such as SASRec \cite{kang2018self}, without intervening the model architecture and the training process.

\noindent $\bullet$ \textbf{Sequential Recommendation Model.} The existing sequential recommendation models encode user representations at a given timestamp based on the interaction sequence:
\begin{equation}\label{eq:encode}
    \begin{aligned}
    \hat{\mathbf{e}}_u^{(t)} = \text{ENC}(\mathbf{s}_{u}^{t}),
\end{aligned}
\end{equation}
where $\mathbf{s}_{u}^{t} = [\hat{\mathbf{e}}_{i_1}, \hat{\mathbf{e}}_{i_2}, \dots \hat{\mathbf{e}}_{i_t}]$ are the embeddings of the items in the interaction sequence. 
All model parameters are trained over historical interaction sequences. 
As to making recommendation, a similarity metric such as inner product is applied to predict the interaction probability between $u$ and $i$ at timestamp $t$: $\hat{p}_{ui}^{t}=f(\hat{\mathbf{e}}_u^{(t)}, \hat{\mathbf{e}}_i)$.

\noindent $\bullet$ \textbf{Post-processing Strategy.} Based on the sequential recommendation model, our goal is to devise a post-processing strategy that recommends items with both high interaction probability and guiding value at each time-step of the guiding process. In particular, we formulate the post-processing operation as an IPG score:
\begin{equation}\label{eq:guiding_score}
    \begin{aligned}
    r_{uij}^{t} &= \hat{p}_{ui}^{t} \cdot g_{uij}^{t},
\end{aligned}
\end{equation}
where $\hat{p}_{ui}^{t}$ reflects the interaction objective; $g_{uij}^{t}$ is a guiding score reflects the guiding value of item $i$. 
The choice of multiplication is to amplify the impact of $\hat{p}_{ui}$, avoiding assigning high IPG scores to items with low interaction probability.
IPG adjusts the item ranking in each iteration as a post-processing strategy once the user representation and interaction probability are updated, making it flexible for integration with most existing recommendation models.

\begin{figure}[!t]
\setlength{\abovecaptionskip}{0cm}
\setlength{\belowcaptionskip}{-0.40cm}
\centering
\includegraphics[scale=0.37]{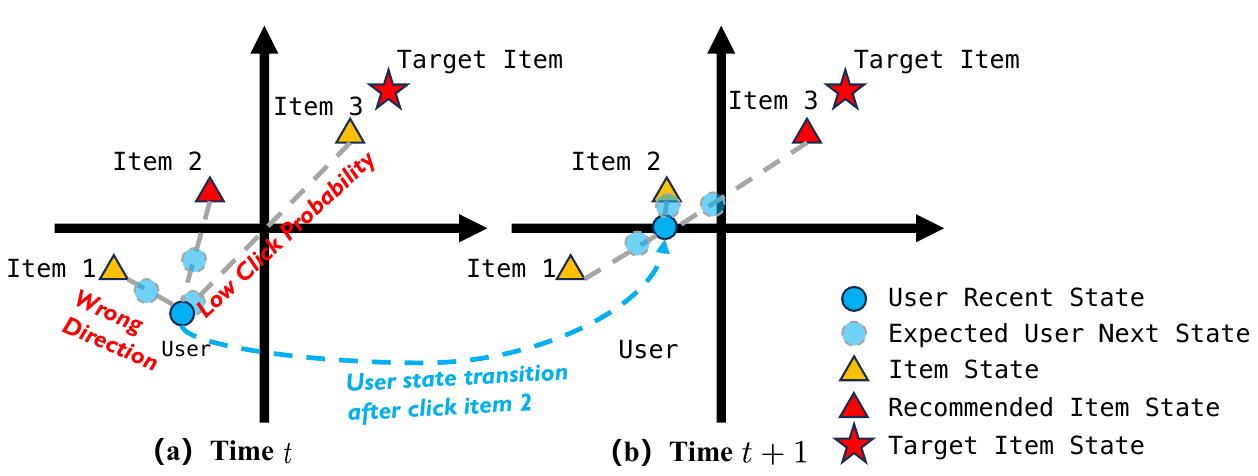}
\caption{Illustration of the proposed IPG framework.}
\label{fig:ill}
\end{figure}

\noindent $\bullet$ \textbf{Guiding Score.} We quantify the increase in user interest towards the target item $j$ after interacting with $i$ at timestamp $t$ as $g_{uij}^{t}$:
\begin{equation}\label{eq:ioi}
    \begin{aligned}
    g_{uij}^{t} &= \hat{\mathbf{e}}_{j}^\top{\hat{\mathbf{e}}}_u^{(t+1)} - \hat{\mathbf{e}}_{j}^\top{\hat{\mathbf{e}}}_u^{(t)},
\end{aligned}
\end{equation}
where $\hat{\mathbf{e}}_u^{(t+1)}$ denotes the user representation after interacting with the recommended item $i$. 
Notably, directly calculating with Eq \eqref{eq:ioi} is time-consuming, as it requires $|\mathcal{I}|$ times of feed-forward process by the sequential encoder to obtain all $\hat{\mathbf{e}}_u^{(t+1)}$ for a single user. 
Considering that obtaining $\hat{\mathbf{e}}_u^{(t+1)}$ mainly involves appending $\hat{\mathbf{e}}_i$ at the end of $\mathbf{s}_u^t$, we make a linear assumption on the change from $\hat{\mathbf{e}}_u^{(t)}$ to $\hat{\mathbf{e}}_u^{(t+1)}$ such that
\begin{equation}
    \hat{\mathbf{e}}_u^{(t+1)} = \gamma \hat{\mathbf{e}}_u^{(t)} + (1 - \gamma) \hat{\mathbf{e}}_i.
\end{equation}
Accordingly, Eq \eqref{eq:ioi} can be simplified as:
\begin{equation}\label{eq:ioi_assumption}
    \begin{aligned}
    g_{uij}^{t} = (1 - \gamma)(\hat{\mathbf{e}}_{i} - {\hat{\mathbf{e}}}_u^{(t)})^\top\hat{\mathbf{e}}_{j}.
\end{aligned}
\end{equation}
Note that $(1 - \gamma)$ can be omitted since it is a constant for all $i$.


The rationale behind IPG is depicted in Figure \ref{fig:ill}. Consider a given target item with three candidate items, IPG evaluates both the guidance direction and interaction probability to derive a recommended item with high guidance efficacy. 
Subsequently, according to Equation \eqref{eq:guiding_score}, IPG selects an item with the highest $r_{uij}^{t}$. For instance, at time $t$, for item 1, despite having a high interaction probability, the user's representation moves further away from the target item, rendering item 1 unfavorable for IPG. On the other hand, although item 3 is the most similar to the target item, the low interaction probability implies recommending it would yield minimal benefits in guiding. 
Consequently, item 2 is selected as the recommendation.

In summary, IPG is a model-agnostic framework that captures real-time user interest through the encoding of the most recent user feedback. Besides, it introduces an explicit IPG score for post-processing, facilitating the ranking of the most valuable items in guiding user preferences toward the target item in each guidance iteration. 
These characteristics render IPG a flexible and effective guiding framework for proactive recommendation.

\section{Experiments}\label{sec:experiment}

\subsection{Environment Simulator}
Conducting offline experiments poses challenges in proactive recommendation tasks, as obtaining real-time groundtruth of users' internal preferences and preference changes towards an item is unrealistic. To address this, we develop an environment simulator to evaluate the effectiveness of our proposed method. 


\textbf{User and item embeddings.} Users and items are characterized by 20-dimensional embeddings, which can be viewed as the concatenation of ten 2-dimensional category embeddings. Consider item $i$ with the embedding $\mathbf{e}_i \triangleq [\mathbf{e}_{i,1}, \dots, \mathbf{e}_{i,10}]$, to construct these embeddings, we start by generating the propensity associated with each category $p_{i,c} \sim \mathcal{U}(0, 1)$. Refered to prior work \cite{DBLP:conf/wsdm/DeffayetTRR23}, we normalize these propensities to $p_{i,c} \leftarrow p_{i,c} / \sum_{c'=1}^{10}p_{i,c'}$. Then each element $e_{i,c,k}$ of the initial embeddings is sampled from $\mathcal{N}(0, 0.4)$ and clamped into $[0, 1]$. The initial category embeddings are multiplied by the corresponding propensity to generate the final embeddings: $\mathbf{e}_{i,c} \leftarrow p_{i,c}\mathbf{e}_{i,c}$, and the embedding is normalized $\|\mathbf{e}_i\| = 1$. We set the category $c$ with the largest $\|\mathbf{e}_{i,c}\|$ as the main category.

\textbf{Click model.} We model the interaction probability between a user $u$ and an item $i$ using a click model: $\sigma\left(w(\mathbf{e}_u^\top\mathbf{e}_i - b_{ui} - b)\right)$, where $\sigma(\cdot)$ denotes the sigmoid function, $b=0.8$ is the bias term, and $b_{ui}$ denotes the item-level boredom effect term.

\textbf{Preference evolution.} Users' preferences are dynamic during the interactions. This phenomenon is reflected by the change of user embeddings in our simulator. After a positive interaction with an item $i$ \cite{DBLP:conf/wsdm/DeffayetTRR23, DBLP:conf/sigecom/DeanM22, DBLP:conf/recsys/CurmeiHRH22}, user $u$'s embedding will move towards the item: $\mathbf{e}_u \leftarrow \gamma \mathbf{e}_u + (1 - \gamma) \mathbf{e}_i$, where $\gamma$ controls the degree of preference evolution and is consistently set to 0.8 for all users.

\textbf{Boredom effect.} Users may feel boredom when they engage with overly similar items frequently. In our simulator, we introduce both category-level boredom effect and item-level boredom effect to reflect this phenomenon.
\begin{itemize}[leftmargin=*]
    \item Category-level: If within the user's last 10 clicked items, 5 or more belong to the same main category $c$, the user's boredom with this category is triggered. Consequently, we adjust the category-embedding as follows: $\mathbf{e}_{u, c} \leftarrow 0.8\cdot\mathbf{e}_{u, c}$. After that, we proceed to re-normalize the user's embedding to ensure that its norm remains equal to 1: $\|\mathbf{e}_u\| = 1$.
    \item Item-level: If a user interacts with the same item excessively, they may gradually lose preference for that particular item while still maintaining an interest in other items of the same category. To account for this, we introduce an item-level boredom effect denoted as a bias term $b_{ui}$ within the click model. This bias term is defined as $b_{ui} = 0.1 \cdot n_{ui}$, where $n_{ui}$ represents the total number of interactions between user $u$ and item $i$.
\end{itemize}

\subsection{Evaluation Protocols}
To evaluate the performance, we propose two phases:
\begin{itemize}[leftmargin=*]
    \item \textbf{Log collection phase: }This phase is designed to gather interaction logs for the offline training of each baseline model. We have specified 100 rounds for this phase, employing a mixed strategy that comprises 30\% oracle recommendations and 70\% random recommendations in each round. Finally, we generate a dataset with 6034 users and 3533 items, with an average of 26.9 positive interactions per user.

    \item \textbf{Guidance phase: }
    This phase entails a 20-round interaction process to guide user preference towards a target item. In this phase, a specific target item is selected as the guiding objective. In each round, recommendation models recommend items to all users, and users subsequently offer feedback.
\end{itemize}
We employ two metrics for the evaluation in guidance phase:
\begin{itemize}[leftmargin=*]
    \item \textbf{Hit Ratio (HR@K):} This metric is to quantify the proportion of items that are interacted positively over K rounds of recommendation. In our settings, we have set K to 20. Formally, $\text{HR@K} = \frac{1}{K|\mathcal{U}|}\sum_{k=1}^{K}\sum_{u\in\mathcal{U}}c_{uk},$
    where $c_{uk} \in \{0, 1\}$ denotes the feedback of $u$ in round $k$.
    \item \textbf{Increase of Interest (IoI@K):} This metric is designed to assess the average increase in user preference towards the target item $j$ following K rounds of recommendations:
    $
    \text{IoI@K} = \frac{1}{|\mathcal{U}|}\sum_{u\in\mathcal{U}}{\mathbf{e}_u^{(K)}}^\top\mathbf{e}_{j} - {\mathbf{e}_u^{(1)}}^\top\mathbf{e}_{j},
    $
    where $\mathbf{e}_u^{(K)}, \mathbf{e}_u^{(1)}$ are the user embeddings in the simulator at round $K$ and the start of the guidance phase respectively.
\end{itemize}

\subsection{Compared Methods}

1) \textbf{SASRec} \cite{kang2018self} is a representative sequential recommender model, which captures the dynamic user preference over time.
2) \textbf{IRN} \cite{DBLP:journals/corr/abs-2211-10002}  is a transformer-based sequential recommendation model that aims to steer users to the target item via extra attention on the target item. 
3) \textbf{SASRec-Heuristic} is a heuristic guiding strategy that introduces an additional term of the inner product between item embedding and target item embedding to obtain the final score of an item to a user:
    $s(u, i, j) = \hat{\mathbf{e}}_u^\top\hat{\mathbf{e}}_i + \alpha \hat{\mathbf{e}}_j^\top\hat{\mathbf{e}}_i$,
where $\alpha$ is a hyper-parameter to control the trade-off between the similarity to the user interest and the similarity to the target item. The approach achieves the guiding objective by recommending items similar to the target item.
4) \textbf{SASRec-IPG} is our proposed method that harnesses a well-trained SASRec and uses Eq \eqref{eq:guiding_score} to generate the recommendation.

\subsection{Results and Discussion}
We randomly select 50 items as the target item and evaluate the performance over 20 rounds of interactions. The results are presented in Table \ref{tab:results}. Notably, SASRec-IPG outperforms all baseline models in terms of the guiding metric IoI@K. The improvement in target preference for IRN is not substantial, potentially attributable to the implicit nature of its guiding objective. Conversely, SASRec-Heuristic emerges as a sub-optimal strategy, as the item-target inner product fails to fully reflect the quality of the item in guiding users to the target item. In contrast, SASRec-IPG demonstrates the ability to guide user preference with an acceptable decline in HR, showcasing its proficiency in accurately capturing the potential increase in user preference towards the target item. 


\begin{table}[t]
\centering
\caption{The overall performance of all methods.}
\vspace{-10pt}
\setlength{\tabcolsep}{0.8mm}{
\resizebox{0.49\textwidth}{!}{
\begin{tabular}{l|ll|ll|ll|ll}
\hline
Model            & HR@5 & IoI@5 & HR@10 & IoI@10 & HR@15 & IoI@15 & HR@20 & IoI@20 \\
\hline
Random           & 0.1058 & 0.0019 & 0.1057 & 0.0042 & 0.1057 & 0.0065 & 0.1056 & 0.0086 \\
SASRec           & 0.5558 & -0.0057 & 0.5552 & -0.0078 & 0.5541 & -0.0063 & 0.5558 & -0.0027      \\
IRN              & 0.5007 & 0.0168 & 0.5011 & 0.0352 & 0.5019 & 0.0510 & 0.5027 & 0.0647 \\
SASRec-Heuristic & 0.3708 & 0.0521 & 0.3687 & 0.1164 & 0.3838 & 0.1683 & 0.3769 & 0.2084      \\
SASRec-IPG       & 0.3992 & 0.0858 & 0.4020 & 0.1707 & 0.4004 & 0.2274 & 0.4019 & 0.2649     \\
\hline
\end{tabular}
}}
\vspace{-0.5cm}
\label{tab:results}
\end{table}

\begin{figure}[t]
\setlength{\abovecaptionskip}{0cm}
\setlength{\belowcaptionskip}{-0.45cm}
\centering
\includegraphics[scale=0.22]{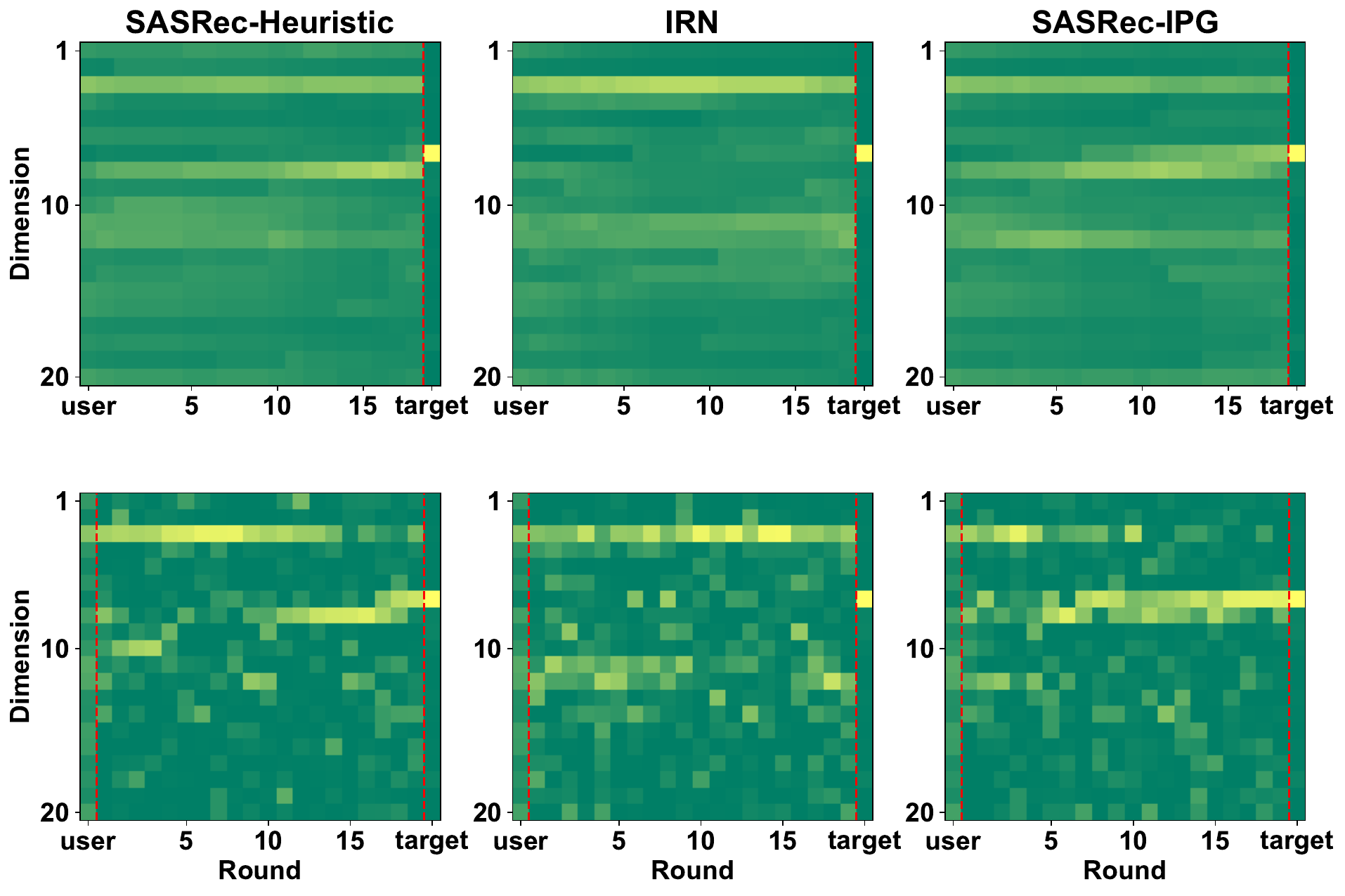}
\caption{The first row shows the embedding evolution of user 61 with target item 3257 and the second row shows the embeddings of recommended items under three methods. The first column of each subfigure is the user's initial embedding and the last column shows the target item's embedding.}
\label{fig:case}
\end{figure}

\textbf{Case Study.} To investigate how SASRec-IPG guides a user's preference towards the target item, we conduct a case study on a user and visualize the ground-truth user embedding evolution in the simulator, and the embeddings of the recommended items, as illustrated in Figure \ref{fig:case}. SASRec-heuristic predominantly caters to the user's initial preference, resulting in relatively static embeddings. A similar phenomenon is observed in IRN. In contrast, SASRec-IPG excels in constructing a guiding trajectory that systematically converges the user's embedding towards the target item. Particularly, it effectively identifies items that encompass both the user's preference features and the target item' features. Such items are regarded as highly efficient in guiding the user's preferences.


\section{Related Work}\label{related_work}
Proactive recommendation is an emerging research domain that primarily comprises two main research lines: 1) Preference shifts in recommendation 2) user preference guiding
. In the domain of preference shifts, prior works predominantly concentrate on understanding how user preferences evolve when interacting with recommender systems, often through simulation-based approaches \cite{DBLP:conf/recsys/CurmeiHRH22, DBLP:conf/sigecom/DeanM22, DBLP:conf/www/PassinoMMAL21}. Some studies also attempt to optimize long-term rewards rather than myopic behaviors under user preference shifts \cite{DBLP:conf/ijcai/IeJWNAWCCB19, DBLP:conf/wsdm/DeffayetTRR23, DBLP:conf/icml/CarrollD0H22}. In terms of preference guiding, early works in dialog systems proposed proactive dialog systems designed to steer conversations towards either a designated target topic \cite{DBLP:conf/acl/TangZXLXH19} or a target topic thread \cite{DBLP:conf/acl/WuGZWZLW19}. Recent studies recommender systems have adopted this concept and introduced a transformer-based sequential influential recommender systems (IRS), which are designed to influence user interest towards a predefined target item \cite{DBLP:journals/corr/abs-2211-10002}. Our work improves the flexibility of IRS and the explicit IPG score design improves the efficacy of guiding compared to IRS. 

\section{Conclusion}\label{conclusion}
We proposed a novel framework IPG, for proactive recommendation. IPG aims to guide users’ preference towards a target item by recommending items that maximize the IPG score that considers
both interaction probability and guiding value. IPG is a model-agnostic post-processing strategy that can be easily integrated with various existing RSs. To evaluate the effectiveness of IPG, we designed an interactive recommendation simulator that simulates users’ preference evolution and feedback. Experimental results showed that IPG can achieve a significant improvement in guiding users’ interests while maintaining a reasonable level of recommendation accuracy. 

In the future, we plan to explore a more universal dynamic evaluation framework. Additionally, leveraging Large Language Models \cite{lin2023multi, DBLP:conf/recsys/BaoZZWF023} for better user understanding and planning more effective guiding paths deserves our investigation. 

\bibliographystyle{ACM-Reference-Format}
\balance
\bibliography{ref.bib}

\newpage
\appendix
\onecolumn
\section{Additional Table and Figure for the Experiments}

\begin{figure}[!h]
     \centering
     \begin{subfigure}[b]{0.43\textwidth}
         \centering
         \includegraphics[width=\textwidth]{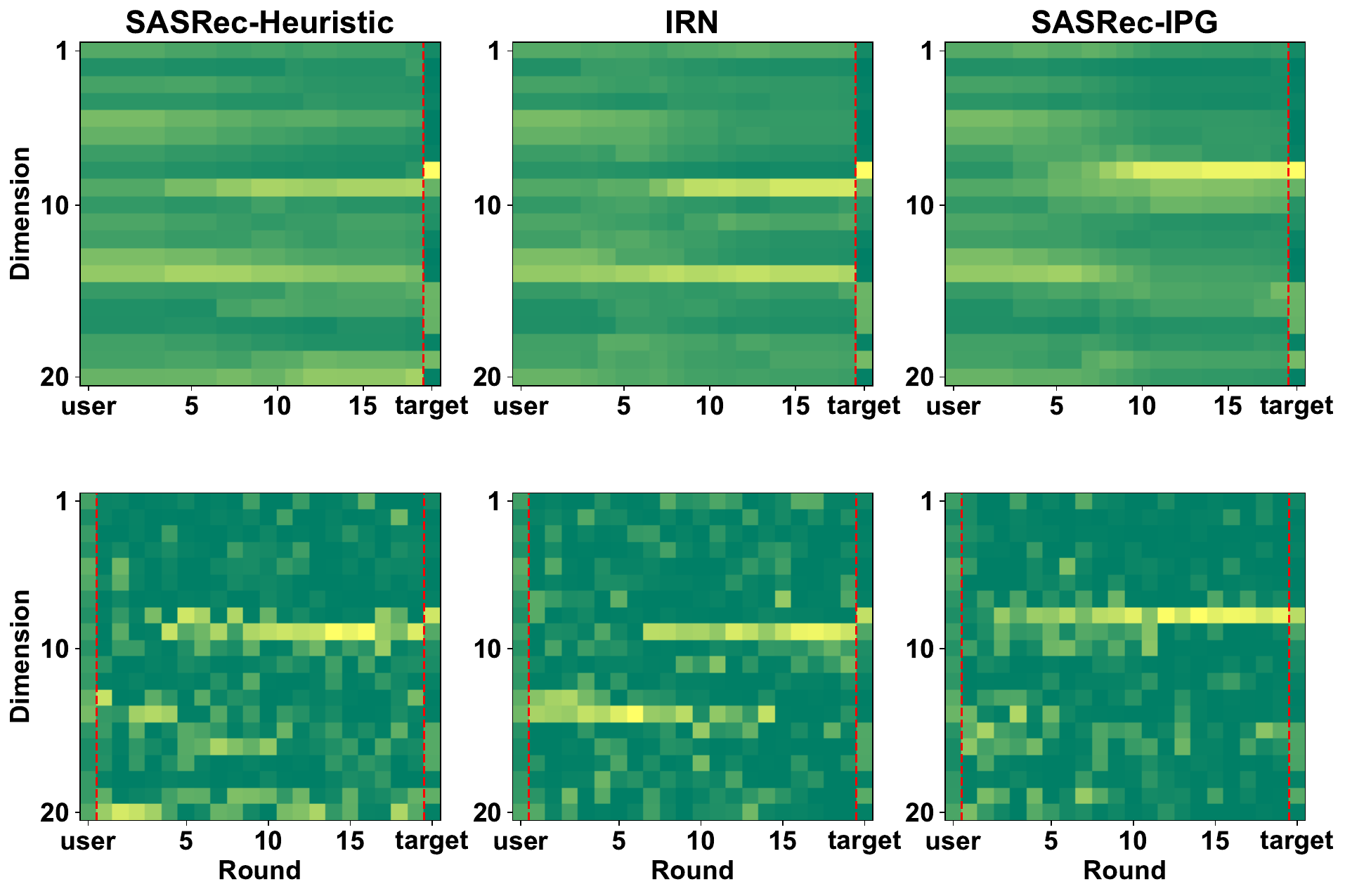}
         \caption{User 2989, target item 1892.}
         \label{fig:a}
     \end{subfigure}
     \begin{subfigure}[b]{0.43\textwidth}
         \centering
         \includegraphics[width=\textwidth]{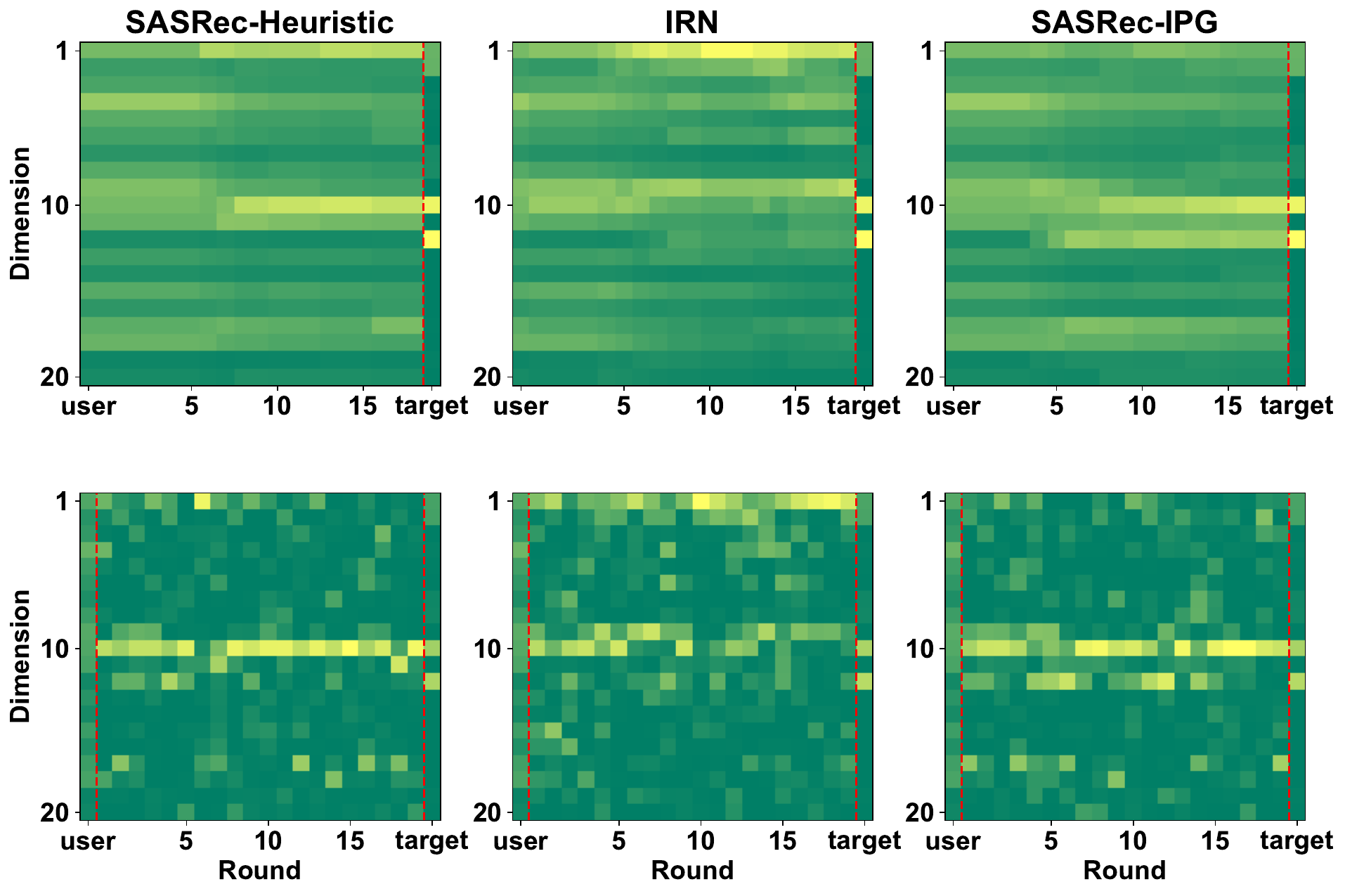}
         \caption{User 5001, target item 865.}
         \label{fig:b}
     \end{subfigure}
     \vfill
     \begin{subfigure}[b]{0.43\textwidth}
         \centering
         \includegraphics[width=\textwidth]{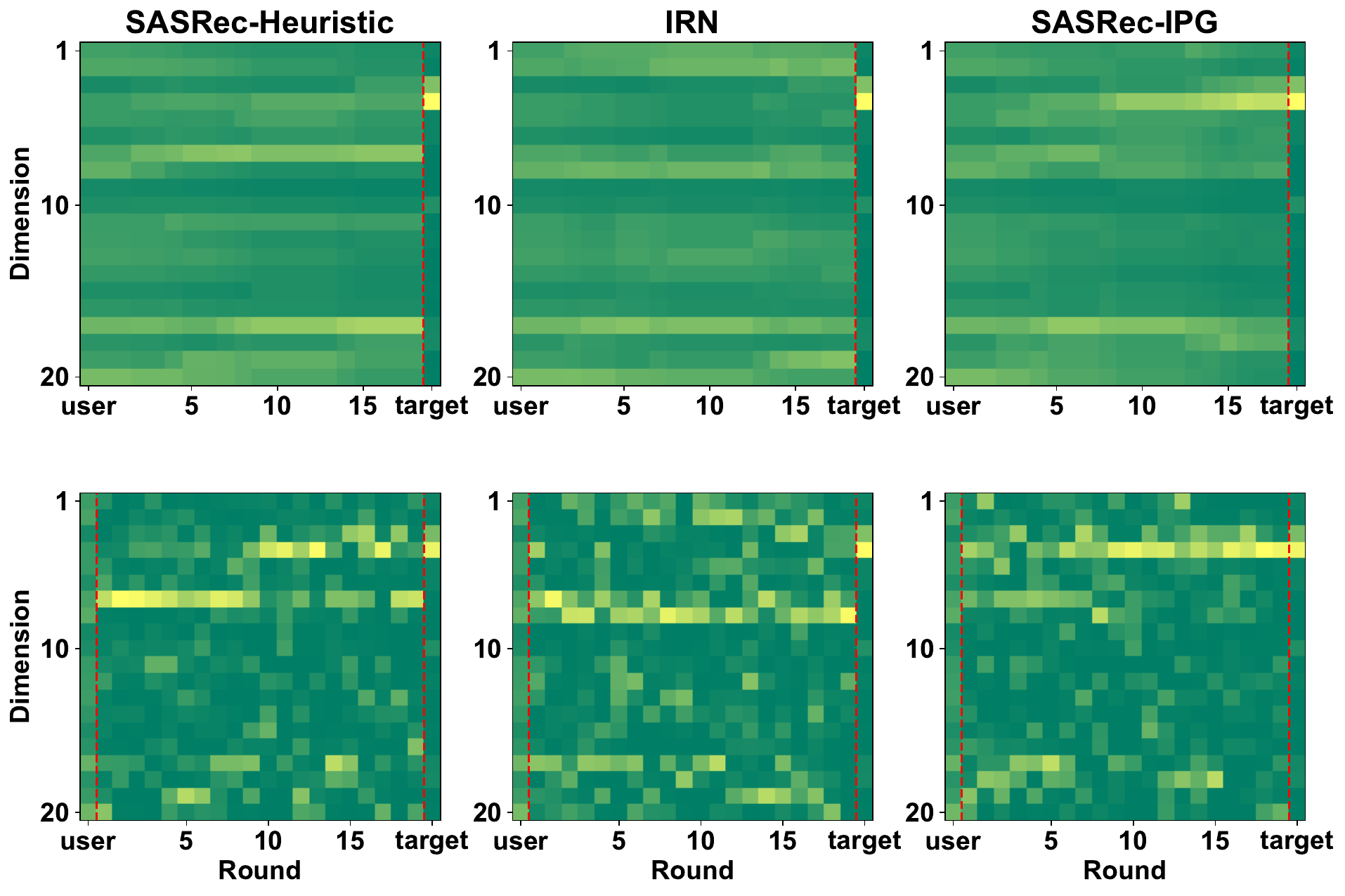}
         \caption{User 377, target item 1327.}
         \label{fig:c}
     \end{subfigure}
     \begin{subfigure}[b]{0.43\textwidth}
         \centering
         \includegraphics[width=\textwidth]{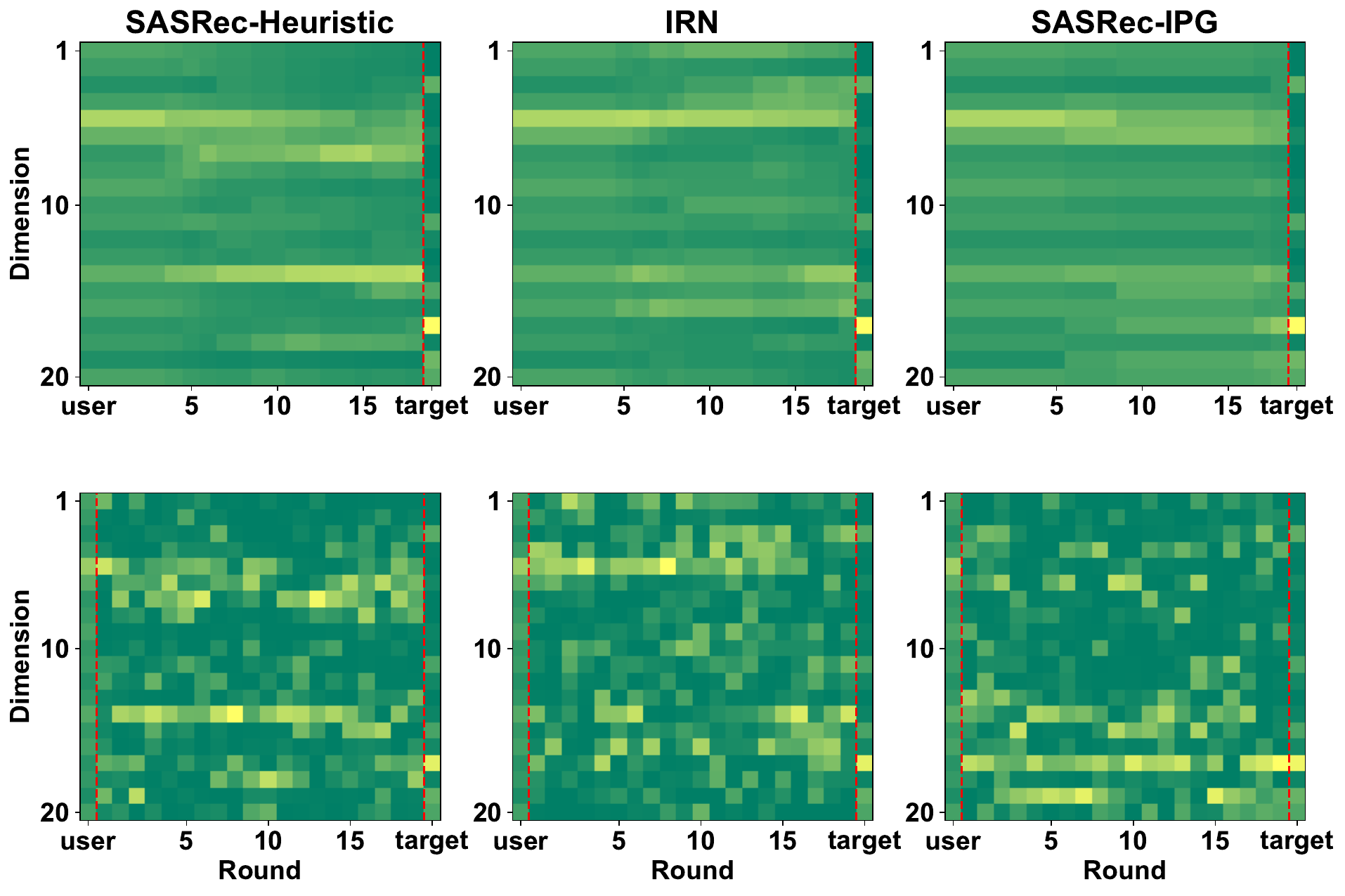}
         \caption{User 6013, target item 2024.}
         \label{fig:d}
     \end{subfigure}
        \caption{Four guiding cases, the logic of each subfigure are in consist with Figure \ref{fig:case}.}
        \label{fig:case_study_all}
\end{figure}

We provide additional results about the experiments in Section \ref{sec:experiment}. Figure \ref{tab:results_all} shows the HR and IoI performance of all methods under different $\gamma$. Figure \ref{fig:case_study_all} demonstrates the embedding evolution and the  embeddings of the recommended items in more guiding cases.

\begin{table*}[!h]
\caption{The overall performance of all methods under different $\gamma$.}

\centering
\begin{tabular}{l|l|ll|ll|ll|ll}
\hline
& Model            & HR@5 & IoI@5 & HR@10 & IoI@10 & HR@15 & IoI@15 & HR@20 & IoI@20 \\
\hline
\multirow{5}{*}{$\gamma=0.6$}
& Random           & 0.0998 & -0.0029 & 0.1000 & -0.0061 & 0.1002 & -0.0089 & 0.1002 & -0.0109 \\
& SASRec           & 0.6248 & -0.0353 & 0.6248 & -0.0487 & 0.6248 & -0.0489 & 0.6248 & -0.0462      \\
& IRN              & 0.5708 & 0.0102 & 0.5702 & 0.0244 & 0.5698 & 0.0382 & 0.5702 & 0.0500 \\
& SASRec-Heuristic & 0.5968 & 0.0904 & 0.5938 & 0.1783 & 0.5993 & 0.2374 & 0.5926 & 0.2748      \\
& \cellcolor{gray!16}SASRec-IPG       & \cellcolor{gray!16}0.5837 & \cellcolor{gray!16}0.1713 & \cellcolor{gray!16}0.5762 & \cellcolor{gray!16}0.3002 & \cellcolor{gray!16}0.5806 & \cellcolor{gray!16}0.3565 & \cellcolor{gray!16}0.5747 & \cellcolor{gray!16}0.3754     \\
\hline
\multirow{5}{*}{$\gamma=0.7$}
& Random           & 0.1037 & 0.0004 & 0.1038 & 0.0008 & 0.1038 & 0.0012 & 0.1038 & 0.0015 \\
& SASRec           & 0.6059 & -0.0195 & 0.6060 & -0.0284 & 0.6060 & -0.0286 & 0.6060 & -0.0254      \\
& IRN              & 0.5527 & 0.0156 & 0.5523 & 0.0337 & 0.5524 & 0.0498 & 0.5525 & 0.0633 \\
& SASRec-Heuristic & 0.5484 & 0.0741 & 0.5455 & 0.1546 & 0.5528 & 0.2147 & 0.5462 & 0.2568      \\
& \cellcolor{gray!16}SASRec-IPG       & \cellcolor{gray!16}0.5214 & \cellcolor{gray!16}0.1329 & \cellcolor{gray!16}0.5182 & \cellcolor{gray!16}0.2522 & \cellcolor{gray!16}0.5206 & \cellcolor{gray!16}0.3194 & \cellcolor{gray!16}0.5171 & \cellcolor{gray!16}0.3539     \\
\hline
\multirow{5}{*}{$\gamma=0.8$}
& Random           & 0.1058 & 0.0019 & 0.1057 & 0.0042 & 0.1057 & 0.0065 & 0.1056 & 0.0086 \\
& SASRec           & 0.5558 & -0.0057 & 0.5552 & -0.0078 & 0.5541 & -0.0063 & 0.5558 & -0.0027      \\
& IRN              & 0.5007 & 0.0168 & 0.5011 & 0.0352 & 0.5019 & 0.0510 & 0.5027 & 0.0647 \\
& SASRec-Heuristic & 0.3708 & 0.0521 & 0.3687 & 0.1164 & 0.3838 & 0.1683 & 0.3769 & 0.2084      \\
& \cellcolor{gray!16}SASRec-IPG       & \cellcolor{gray!16}0.3992 & \cellcolor{gray!16}0.0858 & \cellcolor{gray!16}0.4020 & \cellcolor{gray!16}0.1707 & \cellcolor{gray!16}0.4004 & \cellcolor{gray!16}0.2274 & \cellcolor{gray!16}0.4019 & \cellcolor{gray!16}0.2649     \\
\hline
\end{tabular}

\label{tab:results_all}
\end{table*}

\end{document}